\documentclass[12pt]{iopart}
\usepackage{graphicx}
\usepackage{epsfig}
\providecommand{\openone}{\leavevmode\hbox{\small1\kern-3.8pt\normalsize1}}
\begin{document}

\title{Quantum capacity of dephasing channels with memory}

\author{A D'Arrigo}
\address{MATIS CNR-INFM, Catania \&
Dipartimento di Metodologie Fisiche e Chimiche per l'Ingegneria,
Universit\`a di Catania, Viale Andrea Doria 6, 95125 Catania, Italy} 
\ead{darrigo@femto.dmfci.unict.it}
\author{G Benenti}
\address{CNISM, CNR-INFM \& Center for Nonlinear and Complex Systems, 
Universit\`a degli Studi dell'Insubria, Via Valleggio 11, 22100 Como, Italy}
\address{Istituto Nazionale di Fisica Nucleare, Sezione di Milano,
via Celoria 16, 20133 Milano, Italy}
\ead{giuliano.benenti@uninsubria.it}
\author{G Falci}
\address{MATIS CNR-INFM, Catania \&
Dipartimento di Metodologie Fisiche e Chimiche per l'Ingegneria,
Universit\`a di Catania, Viale Andrea Doria 6, 95125 Catania, Italy} 
\ead{gfalci@dmfci.unict.it}

% 03.67.Hk Quantum communication
% 89.70.+c Information theory and communication theory 
% 03.65.Yz Decoherence; open systems; quantum statistical methods
% 03.67.Pp Quantum error correction and other methods for protection against decoherence

\begin{abstract}
We show that the amount of coherent quantum information that can be 
reliably
transmitted down a dephasing channel with memory is maximized by 
separable input states. In particular, we model the channel as 
a Markov chain or a multimode environment of oscillators. 
While in the first model the maximization is achieved for the 
maximally mixed input state, in the latter it is convenient
to exploit the presence of a decoherence-protected subspace generated
by memory effects. We explicitly compute the quantum channel capacity
for the first model while numerical simulations suggest 
a lower bound for the latter.
In both cases memory effects enhance the coherent information. 
We present results valid for arbitrary size of the input.
\end{abstract}

%\noindent{\it quantum capacity, memory channel, dephasing channel}

\maketitle

\section{Introduction}
Quantum communication channels~\cite{kn:nielsen-chuang,kn:benenti-casati-strini} 
use quantum systems to transfer
classical or quantum information. In the first case, we can encode 
classical bits by means of quantum states. In the latter case, we may
want to transfer an unknown quantum state betweeen differents units
of a quantum system, for instance of a quantum computer, or to 
distribute entanglement between communicating parties. 
In both cases, the fundamental question is what is the maximum rate of 
classical or quantum information that can be faithfully transmitted.
Classical and quantum \textit{capacities}, defined as the maximum 
number of bits/qubits that can be reliably transmitted per channel use,
provide the answer to this question.
 
Quantum channels with memory
are the natural theoretical framework 
for the study of any noisy quantum communication system 
where correlation times are longer than time between consecutive 
uses. This scenario applies 
to optical fibers which may 
show a 
birefringence fluctuating with characteristic time 
longer than the separation between successive 
light pulses~\cite{kn:2004-banaszek-prl} or 
to solid state 
implementations of quantum hardware, 
where memory effects due to
low-frequency impurity noise~\cite{kn:paladino-1}
produce substantial dephasing
\cite{kn:inhom}.

Some theoretical result on quantum 
channels with memory has been already discussed for
transmission  of both 
classical and  
quantum  information through a quantum channel.
With regard to classical information transmission 
down a memory channel, it was pointed out that it 
can be enhanced by using entangled 
input states~\cite{kn:2002-macchiavello-palma-pra,daems,kn:Karimipour}, 
and 
coding theorems 
have been recently proved for classes of memory quantum 
channels~\cite{bjelakovic-boche,datta-dorlas}. 
Concerning quantum capacity, 
a lower bound has been found for some classes 
of channels with memory~\cite{hamada} and subsequently  
specific model environments (structured in two parts, one responsible for 
memory effects and the other acting as a memoryless environment) have been 
studied~\cite{bowenmancini,giovannetti,werner}. In particular, 
coding theorems for quantum capacity have been proved in~\cite{werner}
for the so-called \textit{forgetful channels}, for which memory effects
decay exponentially with time.

The problem is formalized by considering the 
$N$-uses  Hilbert space $\mathcal{H}_N=\mathcal{H}^{\otimes N}$ and 
defining the system ${\textsf S}$, described by the 
reduced density matrix (RDM) $\rho$ for $N$ uses. 
The input state is $\rho=\sum_{i=1}^K p_i \rho_i$, namely
states chosen from the ensemble $\{\rho_1,...,\rho_K\}$, 
with a priori probabilities $\{p_1,...,p_K\}$, are sent down 
the channel.
Due to the coupling to further uncontrollable degrees of 
freedom, the transmission of $\textsf S$ may be noisy.
The output is therefore described by a linear, completely positive, 
trace preserving (CPT) map 
$\mathcal{E}_N(\rho)$, corresponding to $N$-uses (the single use is defined in 
$\mathcal{H}$ and described by $\mathcal{E}$).
The map $\mathcal{E}_N(\rho)$ 
can always be represented starting from an enlarged 
vector space including a suitably chosen 
environment $\textsf E$, initially in a pure state: 
$w_0 \equiv |0\rangle _{\tiny \textsf E}\!\langle 0|$
\begin{equation}
%&&
\mathcal{E}_N(\rho) =
\mathrm{Tr}_{\tiny \textsf E}[U \, (\rho\otimes w_0) \,U^\dagger],
\label{eq:map}
\end{equation}
where $U$ is a suitable unitary 
evolution of $\textsf{S+E}$ referring to $N$ uses. 
The conditional 
(depending on $\rho$) evolution of 
the environment can also be considered. It is 
described by the environment RDM and allows to define 
the conjugate CPT map,
$w=
\mathrm{Tr}_{\tiny \textsf S}[U\, (\rho \otimes w_0)\, U^\dagger]
=: \tilde{\mathcal{E}}_N(\rho)$.

The \textit{quantum capacity} $Q$ refers to the coherent 
transmission of quantum information (measured
in number of qubits), and it is related to the 
dimension of the largest subspace of $\mathcal{H}_N$
reliably transmitted down the channel, in the limit of large $N$. 
The value of $Q$ can be computed, for memoryless channels, as 
\cite{lloyd,barnum,shor,devetak,winter}
\begin{eqnarray}
&&Q \,=\, \lim_{N\to\infty} \frac{Q_N}{N},
\quad \quad 
Q_N\,=\,\max_{\rho} I_c(\mathcal{E}_N,\rho),
\label{qinfo}
\\&&
I_c(\mathcal{E}_N,\rho)
\,=\,S[\mathcal{E}_N(\rho)]-
S_e^N(\rho).
\label{coinfo}
\end{eqnarray}
Here 
$S(\rho)=-\mathrm{Tr}[\rho \log_2 \rho]$ is the von 
Neumann entropy, $S_e^N(\rho) \equiv 
S[\tilde{\mathcal{E}}_N(\rho)]$ is the  
\textit{entropy exchange} \cite{schumacher}. The quantity 
$I_c(\mathcal{E}_N,\rho)$ is called  
\textit{coherent information} \cite{schumachernielsen}
and must be maximized  
over {\em all input states} $\rho$.

The limit $N\to\infty$ in (\ref{qinfo}) 
makes difficult the evaluation of $Q$. On the other
hand this regularization is necessary, 
since in general $I_c$ is not 
subadditive. Indeed
for entangled input states $\rho$ \cite{barnum}
we may have 
$I_c(\mathcal{E}_N,\rho)>\sum_{k=1}^N 
I_c(\mathcal{E},\rho^{(k)})$, where 
$\rho^{(k)}=\mathrm{Tr}_{{\scriptsize \textsf S} - (k)}(\rho)$ refers to 
the individual transmission of the $k-$th unit of 
information,
therefore in general it cannot be excluded that 
${Q_N}/{N} > Q_1$. 
The regularization is not necessary if 
the final state
$w$ of $\textsf E$ can be reconstructed from
the final state $\rho^\prime$ of the system. In this case,
referred to as \textit{degradable 
channels} \cite{devetakshor, CarusoGiovannetti, CarusoGiovannettiHolevo,
WolfPerezGarcia}, 
it exists
a CPT map $\mathcal{T}$ such that 
$\tilde{\mathcal{E}}=\mathcal{T}\circ \mathcal{E}$. 
It turns out \cite{devetakshor} that for degradable 
channels the coherent 
information $I_c(\mathcal{E}_N,\rho)$ reduces to a suitable 
conditional entropy~\cite{kn:nielsen-chuang}, 
which is subadditive and concave in the input state $\rho$, 
and therefore the quantum capacity
is given by the ``single-letter'' formula 
$Q=Q_1$.

In this work we focus on dephasing channels with memory.
Dephasing channels 
are characterized by the property that  
when $N$ qubits are sent through the 
channel, the states of a preferential 
orthonormal basis $\{|j\rangle \equiv 
|j_1,....,j_N\rangle, \,j_1,...,j_N=0,1\}$
are transmitted without errors, implying a conservation 
law to hold~\cite{kn:conservations}. 
Therefore, dephasing channels are noiseless from the viewpoint
of the transmission of classical information, since 
the states of the preferential basis can be used
for encoding classical information.
Of course {\em superpositions} of basis states may decohere,
thus corrupting the transmission of quantum information.
Dephasing channels are relevant for systems 
in which relaxation is much slower than 
dephasing~\cite{chuang,kn:paladino-1}. 
When memory effects are taken into account, we have that 
${\cal E}_N \neq {\cal E}^{\otimes N}$, i.e. 
the channel does not act on each carrier {\it independently}.

We show that the coherent
information is maximized by  input states separable
and diagonal in the reference basis $\{|j\rangle\}$. 
In particular, we calculate the coherent information for two models of 
dephasing channels. 
For a Markov chain we show that the coherent
information is maximized by maximally mixed input states and
compute $Q$. For an environment modeled by a bosonic bath,
we propose a coding strategy based on the existence of a
decoherence-protected subspace generated by memory effects
and use numerical results to suggest a lower bound for 
$Q$. It turns out that in both cases memory effects increase 
the coherent information.

\section{The dephasing channel and quantum capacity}
The unitary representation of 
the generalized dephasing channel~\cite{devetakshor} reads 
\begin{equation}
U |{ j\rangle |0_{\tiny \textsf E}\rangle = 
|j\rangle|\phi_j}\rangle,
\label{Ugendeph}
\end{equation}
where $|\phi_{j}\rangle$ are 
environment states, 
in general non mutually orthogonal, 
describing the conditional evolution.
The map $\mathcal{E}_N$ can be written in the Kraus 
representation~\cite{kn:nielsen-chuang,kn:benenti-casati-strini} as 
\begin{equation}
\rho^\prime = \mathcal{E}_N(\rho) = 
\sum_\alpha\, A_\alpha\, \rho\, A_\alpha^\dagger,
\label{generalizedkraus}
\end{equation}
where the system operators 
$(A_\alpha)_{jl}=\langle \alpha_{\tiny \textsf E} | \phi_{j}\rangle \,
\delta_{jl}$
are diagonal in the reference basis 
(here $\{|\alpha_{\tiny \textsf E}\rangle\}$ is an
orthonormal basis for the environment).
It is easily shown that this channel is 
degradable~\cite{devetakshor}. 
Indeed, for a generic input 
$\rho=\sum_{j,l} \rho_{jl}  |j\rangle\langle l|$, 
equation (\ref{Ugendeph}) yields
\begin{equation}
w=\tilde{\mathcal{E}}_N(\rho)=\sum_j \rho_{jj}
|\phi_j\rangle \langle\phi_j|.
\end{equation} 
Since $w$ only depends on 
the populations $\rho_{jj}$ which are conserved, we 
can write as well 
$\tilde{\mathcal{E}}_N=
\tilde{\mathcal{E}}_N\circ\mathcal{E}$,
thus proving degradability.

We now show that for 
a generalized dephasing channel
the coherent information 
$I_c(\mathcal{E}_N,\rho)$ 
is maximized 
by input states
diagonal in the reference basis. 
To this end we introduce
\begin{equation}
\rho_k=\mbox{$\frac{1}{2}$}\,
\big(\rho_{k-1}+\Sigma_z^{(k)}\rho_{k-1}\Sigma_z^{(k)}
\big)\;,
\label{sigmaziteration}
\;\;\;(k=1,...,N),
\end{equation}
where $\rho_0=\rho$ and the local operator 
$\Sigma_z^{(k)}=\openone^{(1)}\otimes\cdots\otimes
\openone^{(k-1)}\otimes \sigma_z^{(k)}\otimes \openone^{(k+1)}
\otimes\cdots\otimes \openone^{(N)}$ acts non-trivially only on the 
$k-$th qubit, by the Pauli operator $\sigma_z^{(k)}$ 
which has eigenvectors ${|j_k\rangle}$.
We can easily see that 
$\rho_N$ is the diagonal part of $\rho$, by using the
standard representation of the $N$-qubit density matrix: 
\begin{equation}
\rho=\sum_{\{i_k\}} c_{i_1...i_N} 
\sigma_{i_1}^{(1)}\otimes \cdots \otimes \sigma_{i_N}^{(N)},
\quad  \quad i_k=0,x,y,z,
\end{equation}
where $\sigma_0=\openone$. We now study the action of the 
operators $\Sigma_z^{(k)}$. First of all 
$\mathcal{E}_N(\Sigma_z^{(k)}\rho\Sigma_z^{(k)})=
\Sigma_z^{(k)}\mathcal{E}_N(\rho)\Sigma_z^{(k)}$ for any 
$k$ and $\rho$, 
since $\Sigma_z^{(k)}$ commutes with 
the Kraus operators in (\ref{generalizedkraus}). Also
$S[\Sigma_z^{(k)}\mathcal{E}_N(\rho)\Sigma_z^{(k)}]=
S[\mathcal{E}_N(\rho)]$,
since the von Neumann entropy is invariant under unitary 
local transformations. Moreover
$\tilde{\mathcal{E}}_N(\Sigma_z^{(k)}\rho\Sigma_z^{(k)})=
\tilde{\mathcal{E}}_N(\rho)$, since the populations of 
$\Sigma_z^{(k)}\rho\Sigma_z^{(k)}$ 
are the same as for $\rho$. We can therefore conclude that
$I_c(\mathcal{E}_N,\Sigma_z^{(k)}\rho\Sigma_z^{(k)})=I_c(\mathcal{E}_N,\rho)$.
This latter relation, together with the concavity of the coherent 
information for degradable channels (a direct consequence of 
the concavity of the conditional von Neumann entropy) implies that
\begin{equation}
I_c(\mathcal{E}_N,\rho_N)\ge 
I_c(\mathcal{E}_N,\rho_{N-1})\ge\cdots\ge 
I_c(\mathcal{E}_N,\rho_0).
\label{Iconcave}
\end{equation}
Hence, diagonal input states maximize the coherent information. 
These states are separable, since they can be written in the form
\begin{equation}
\rho_N=\sum_{j_1,...,j_N}q_{j_1...j_N} 
\rho_{j_1}^{(1)}\otimes \cdots \otimes
\rho_{j_N}^{(N)},
\end{equation}
with $\rho_{j_k}^{(k)}\equiv |j_k\rangle\langle j_k|$,
($k=1,...,N$), $0\leq q_{j_1...j_N} \leq 1$ and 
$\sum_{j_1,...,j_N} q_{j_1...j_N}=1$.

\section{The memory dephasing channel}
\subsection{Forgetful channels}
Interesting results on the quantum capacity of dephasing channels 
with memory can be obtained for forgetful channels, for which 
the memory dies out exponentially with time.
Forgetfulness is defined in~\cite{werner}, according
to a model in which the environment 
is structured in two parts: a memoryless one
and one responsible for memory effects
(see also ~\cite{bowenmancini}).
A key feature of forgetfulness is that it permits,
with a negligible error, the mapping of the memory channel itself 
into a memoryless one. 
This may be clarified by referring to the 
double-blocking strategy~\cite{werner}:
we consider blocks of $N+L$ uses of the channel and  
do the actual coding and decoding for the first $N$ uses, ignoring the 
remaining $L$ idle uses. The resulting CPT map $\bar{\mathcal{E}}_{N+L}$
acts on density matrices $\rho$ on $\mathcal{H}^{\otimes N}$.
If we consider $M$ uses of such blocks, the corresponding CPT map
$\bar{\mathcal{E}}_{M(N+L)}$ can be approximated by
the memoryless setting $(\bar{\mathcal{E}}_{(N+L)})^{\otimes M}$.
This is possible because correlations among different blocks 
decay during the idle uses. This property can be expressed 
as follows\cite{werner}: 
\begin{equation}
\Vert \bar{\mathcal{E}}_{M(N+L)}(\rho) -
(\bar{\mathcal{E}}_{(N+L)})^{\otimes M}(\rho) \Vert_1
\leq h\,(M-1) c^{-L},
\label{eq:forgetful}
\end{equation}
for any input state $\rho$ in $\mathcal{H}^{\otimes MN}$,
where $c>1$,  $\Vert  \cdot  \Vert_1$ is 
the trace distance~\cite{kn:nielsen-chuang}, 
and $h$ is some constant depending on the memory 
model (note that $c$ and $h$ 
are independent of the input state).
This equation states that, even though the error commited by replacing
the memory channel itself with the corresponding memoryless channel grows 
with the number $M$ of blocks, it goes to zero expontially fast with 
the number $L$ of idle uses in a single block.
Equation (\ref{eq:forgetful}) permits the
proof of coding theorems for forgetful quantum memory 
channels, by mapping them into the corresponding
memoryless channels, for which quantum coding theorems hold~\cite{werner}.
In particular, the quantum capacity $Q$ is $\lim_{N\to\infty} Q_N/N$. 
Equation (\ref{eq:forgetful}) by itself is a sufficient condition to prove
coding theorems. Therefore, in the following we will use the wording 
forgetful channel for any system satisfying inequality (11), 
independently of the model from which memory arises.
Now we focus on two specific, physically 
significant models.

\subsection{Markovian model}
The first model is a 
quantum channel that maps an arbitrary 
$N$-qubit input state $\rho$ onto 
\begin{equation}
\rho^\prime= \mathcal{E}_N(\rho)=\sum_{i_1,...,i_N} 
A_{i_1...i_N} \rho A^\dagger_{i_1...i_N},
\quad \quad i_k=0,z,
\label{dephmemory}
\end{equation}
where the Kraus operators $A_{i_1...i_N}$ are defined in
terms of the Pauli operators $\sigma_0= \openone$ and
$\sigma_z$: 
\begin{equation}
A_{i_1...i_N}=\sqrt{p_{i_1...i_N}}B_{i_1...i_N},\quad
B_{i_1...i_N}\equiv\sigma_{i_1}^{(1)}\otimes
\cdots\otimes\sigma_{i_N}^{(N)},
\label{Krausnuses}
\end{equation}
with $\sum_{\{i_k\}} p_{i_1...i_N}=1$ and 
$\sigma_{i_k}^{(k)}$ acting on the $k$-th qubit\footnote{The 
Kraus operators (\ref{Krausnuses}) define a generalized 
dephasing channel in the sense of equation (\ref{Ugendeph}),
with 
$$
U=\sum_{i_1,...,i_N} \sqrt{p_{i_1...i_N}} 
\sigma_{i_1}^{(1)}\otimes
\cdots \otimes \sigma_{i_N}^{(N)}\otimes 
|i_1...i_N\rangle_{\tiny \textsf E}\langle
0...0|.
$$}.

The quantity $p_{i_1...i_N}$ can be interpreted as the 
probability that the ordered sequence 
$\sigma_{i_1}^{(1)},...,\sigma_{i_N}^{(N)}$ 
of Pauli operators is applied to the $N$ qubits crossing the channel.
We define the single-qubit marginal probability 
$p_{i_q}=\sum_{\{i_k, k\neq q\}}
 p_{i_1...i_N}$
and similarly the two-qubit marginal probability $p_{i_{q^\prime}i_q}$
and assume that $\{p_{i_q}\}=\{1-p_z,p_z\}$ for all $q=1,\dots,N$. 
Under these conditions the maximum of coherent information in model 
(\ref{dephmemory}) is obtained 
for the totally unpolarized
input state 
$\rho_{unp}\equiv\ (1/2^N) \openone^{\otimes N}$.
To prove this statement, we construct the same iterative 
transformation 
as in (\ref{sigmaziteration}) but with 
$\Sigma_x^{(k)} 
=\openone^{(1)}\otimes\cdots\otimes \openone^{(k-1)}
\otimes \sigma_x^{(k)}\otimes \openone^{(k+1)} 
\otimes\cdots\otimes \openone^{(N)}$ 
instead of $\Sigma_z^{(k)}$,
and notice that 
$\rho_N=\rho_{unp}$ is obtained 
starting from an 
input state $\rho_0$ diagonal in the reference basis. 
Moreover it can be proven that in this case 
$$
S[\mathcal{E}_N(\Sigma_x^{(k)}\rho_0 \Sigma_x^{(k)})]=
S[\Sigma_x^{(k)}\rho_0\Sigma_x^{(k)}]=S(\rho_0).
$$
Since $\rho_0$ is diagonal and $\mathcal{E}_N$ only changes
off-diagonal matrix elements, then $\mathcal{E}_N(\rho_0)=\rho_0$
and $S[\mathcal{E}_N(\Sigma_x^{(k)}\rho_0 \Sigma_x^{(k)})]=
S[\mathcal{E}_N(\rho_0)]$. We can also prove that
$$
S[\tilde{\mathcal{E}}_N(\Sigma_x^{(k)}\rho_0\Sigma_x^{(k)})]=
S[\tilde{\Sigma}_z^{(k)}\tilde{\mathcal{E}}_N(\rho_0)
\tilde{\Sigma}_z^{(k)}]=S[\tilde{\mathcal{E}}_N(\rho_0)]. 
$$
Here $\tilde{\Sigma}_z$ is defined as $\Sigma_z$ but acts on the
environment.
Therefore, 
$I_c(\mathcal{E}_N,\Sigma_x^{(k)}\rho_0\Sigma_x^{(k)})=
I_c(\mathcal{E}_N,\rho_{0})$. Taking again advantage of the concavity 
of coherent information for degradable channels, we 
finally obtain
\begin{equation}
I_c(\mathcal{E}_N,\rho_{unp})\ge 
I_c(\mathcal{E}_N,\rho_{0}).
\end{equation}

We can explicitly compute the quantum capacity 
when the joint probabilities in 
equation (\ref{Krausnuses})
are described by a Markov 
chain~\cite{kn:2002-macchiavello-palma-pra,hamada}: 
\begin{equation}
p_{i_1,...,i_N}=p_{i_1}p_{i_2|i_1}\cdots p_{i_N|i_N-1},
\label{eq:markov}
\end{equation}
where 
\begin{equation}
p_{i_k|i_{k-1}}=(1-\mu)\,p_{i_k}+\mu\,\delta_{i_k,i_{k-1}}.
\label{eq:propagator}
\end{equation}
Here $\mu\in[0,1]$ measures the partial 
memory of the channel: it is the  probability that 
the same operator 
(either $\openone$ or $\sigma_z$) 
is applied for two consecutive uses
of the channel, whereas $1-\mu$ is the probability that 
the two operators are uncorrelated.
The limiting cases $\mu=0$ and 
$\mu=1$ 
correspond to memoryless channels and channels with 
perfect memory, respectively. In this noise model $\mu$ might depend on
the time interval between two consecutive channel uses. 
If the two qubits are sent at a time interval $\tau \ll \tau_c$,
where $\tau_c$ denotes the characteristic memory time scale for the environment,
then the same operator is applied to both qubits ($\mu=1$), 
while the opposite limit corresponds to
the memoryless case ($\mu=0$).

The Markov chain model is forgetful, since condition
(\ref{eq:forgetful}) is fulfilled. 
We first consider a sequence of two blocks of $N+L$ 
channel uses, for which 
\begin{equation}
\rho^\prime =
\bar{\mathcal{E}}_{2(N+L)}(\rho)=\sum_{I} p_I B_I \rho B_I^\dagger,
\end{equation}
where the index $I$ stands for $i_1,...,i_N,i_{N+L+1},...,i_{2N+L}$
and the operators $B_I$ are defined in equation (\ref{Krausnuses}). 
The output state $\rho^\prime$ can be approximated by
\begin{equation}
\tilde{\rho}^\prime =
(\bar{\mathcal{E}}_{N+L})^{\otimes 2}(\rho)=
\sum_{I} \tilde{p}_I B_I \rho B_I^\dagger,
\end{equation}
where the factorized probability distribution
$\tilde{p}_I\equiv p_{i_1,...,i_N}p_{i_{N+L+1},...,i_{2N+L}}$.
Taking advantage of the strong convexity of trace distance~\cite{kn:nielsen-chuang},
we obtain
\begin{equation}
\Vert \rho^\prime - \tilde{\rho}^\prime \Vert_1 \leq 
D(p_I,\tilde{p}_I),
\end{equation}
where the Kolmogorov distance between the probability distributions 
$\{p_I\}$ and $\{\tilde{p}_I\}$ is defined as 
\begin{equation}
D(p_I,\tilde{p}_I)=\frac{1}{2}\sum_I |p_I-\tilde{p}_I|.
\end{equation}
Using the properties of stationary Markov chains and equation (\ref{eq:propagator})
we obtain
\begin{equation}
D(p_I,\tilde{p}_I)\leq 2 \mu^{L+1}.
\end{equation}
This implies
\begin{equation}
\Vert \bar{\mathcal{E}}_{2(N+L)}(\rho) -
(\bar{\mathcal{E}}_{(N+L)})^{\otimes 2}(\rho) \Vert_1
\leq 2\mu^{L+1},
\end{equation}
from which equation (\ref{eq:forgetful}) readily follows\footnote{It 
is interesting to remind the reader that the Markov chain model can also be 
formulated in terms of a structured environment~\cite{bowenmancini,werner}.}.
The forgetfulness of the Markov chain model allows us to compute the 
quantum capacity from the regularized coherent information (\ref{qinfo})
\cite{werner}.

In order to compute the quantum capacity, we consider the input
state $\rho_{unp}$ and evaluate the coherent information
$I_c(\mathcal{E}_N,\rho_{unp})$. 
In this case $S[\mathcal{E}_N(\rho_{unp})]=
S(\rho_{unp})=N$. 
We now take advantage of the formula 
$(S_e)_N=S(W)$, where the density operator $W$
has components $W_{i_1...i_N,i_1'...i_N'}=
\mathrm{Tr}(A_{i_1...i_N}\, \rho\, A_{i_1'...i_N'}^\dagger)$ 
\cite{schumacher}. Here $W$ is diagonal and 
\begin{equation}
S(W)=-\sum_{\{i_k\}}p_{\{i_k\}}\log_2 p_{\{i_k\}}
\equiv H(X_1,...,X_N),
\end{equation}
where $H(X_1,...,X_N)$ is by definition the Shannon entropy 
of the collection of random variables $X_1,...,X_N$
(characterized by the joint probabilities $p_{i_1...i_N}$).
For a stationary Markov chain, we have~\cite{coverthomas}
$$%\begin{equation}
\lim_{N\to\infty}\frac{1}{N}H(X_1,...,X_N)=H(X_2|X_1)=
p_0 H(q_0) + p_z H(q_z),
$$%\end{equation}
where $q_{0,z} \equiv (1-\mu)p_{0,z}+\mu$ are the conditional 
probabilities that the channel acts on two subsequent qubits via
the same Pauli operator,
and $H(q_0)$, $H(q_z)$ are binary Shannon entropies,
defined by $H(q)=-q\log_2 q -(1-q)\log_2(1-q)$. 
Therefore, the quantum capacity is given by
\begin{equation}
Q=1-p_0H(q_0)-p_zH(q_z).
\label{capacitymarkov}
\end{equation}
It is interesting to point out that $Q$ 
increases for increasing 
degree of memory of the channel. 
In particular, for $\mu=0$ we 
recover the capacity $Q=Q_1=1-H(p_0)$ of the memoryless dephasing 
channel, while for perfect memory ($\mu=1$) 
$Q=1$, that is, the channel is asymptotically noiseless
\cite{bowenmancini}. 
We also note that the right hand side of 
~(\ref{capacitymarkov}) is known 
\cite{hamada} to be a lower bound for the quantum capacity
of the Markov chain dephasing channel. Our results prove that
this bound is tight.

\begin{figure}
\centering
\includegraphics[width=80mm,height=53mm]{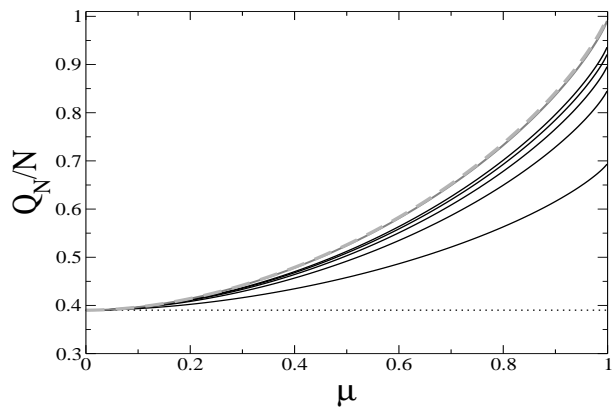}
\caption{Plot of $Q_N/N$ as a function of $\mu$,
for the Markov chain model (\ref{eq:markov}), with $p_0=0.85$.
From bottom to top: $N=2,4,6,8,10$ (black curves), 
$N=100,\infty$ (grey curves).
The dotted black line gives the memoryless quantum capacity.}
\label{fig:markov}
\end{figure}

In order to illustrate the convergence of $Q_N/N$ to its limiting value
$Q$, we first compute the entropy exchange for the $N$-qubit input state 
$\rho_{unp}$. It is easy to check that 
\begin{equation}
(S_e)_N=p_0H(q_0)+p_zH(q_z)+(S_e)_{N-1}.
\end{equation}
Using this recurrence relation we obtain
\begin{equation}
(S_e)_N=(N-1)[p_0H(q_0)+p_zH(q_z)]+(S_e)_1,
\end{equation}
where $(S_e)_1=H(p_0)$. Therefore
\begin{equation}
Q_N=N-(N-1)[p_0H(q_0)+p_zH(q_z)]-H(p_0).
\end{equation}
A plot of $Q_N/N$ for various $N$ as a function of the memory 
factor $\mu$ is shown in figure~\ref{fig:markov}.
It is clear that the convergence of $Q_N/N$ is faster when
the memory factor is smaller. Indeed, it is easy to prove
that
\begin{equation}
\epsilon_N\equiv
Q-\frac{Q_N}{N}
\end{equation}
is a growing function of $\mu$, with $\epsilon_N(\mu=0)=0$
and $\epsilon_N(\mu=1)=H(p_0)/N$. Moreover, for $\mu\ll 1$
we obtain
\begin{equation}
\epsilon_N(\mu)\approx
\frac{1}{2\ln 2}\frac{\mu^2}{N}.
\end{equation}

\subsection{Spin-boson model}
The second model of dephasing channel is defined by 
the system (qubits)-environment Hamiltonian
\begin{equation}
\label{bosonicbath}
 {H}(t) = H_{\tiny \textsf E} -\frac{1}{2}{X}_{\tiny \textsf E}{F}(t) + H_C.
\end{equation}
Here ${H}_{\tiny \textsf E} =
\sum_\alpha \omega_\alpha {b}^\dagger_\alpha {b}_\alpha$
is a bosonic bath and 
${X}_{\tiny \textsf E}=\sum_\alpha  
({b}^\dagger_\alpha + {b}_\alpha)$ is the environment 
operator coupled to the qubits. 
The $k$-th qubit has a switchable 
coupling to the environment via its Pauli operator 
$\sigma_z^{(k)}$:
\begin{equation}
{F}(t)=\lambda
          \sum_{k=1}^{N}\sigma_z^{(k)}\, f_k(t),
\end{equation}
where $f_k(t)=1$ when the qubit is inside the channel, 
and $f_k(t)=0$ otherwise. Finally,
\begin{equation}
H_C=\sum_\alpha \frac{\lambda^2}{4\omega_\alpha}
\sum_{k=1}^N \sigma_z^{(k)} 
\end{equation}
is a counterterm \cite{weiss}.
We call $\tau_p$ the time each
carrier takes to cross the channel and $\tau$ the
time interval that separates
two consecutive qubits entering the channel. 
The Hamiltonian (\ref{bosonicbath}) is expressed in 
the interaction
picture with respect to the qubits. 
If initially the system and the environment are
not entangled, the state of the system at time $t$ 
is given by the map (\ref{eq:map}) where
\begin{equation} \label{evolution_operator_01}
U(t) = {T} e^{-\frac{i}{\hbar} \int_{0}^t ds H(s)}.
\end{equation}
In particular, we are interested in the final state 
$\rho^\prime=\rho(t=\tau_N)$, where 
$\tau_N=\tau_p+(N-1)\tau$ 
is the transit time across the channel for the $N$-qubit train.
To treat this problem we choose the factorized basis states
$\{|j \, \alpha_{\tiny \textsf E}\rangle\}$, where - as above - 
$\{|j \rangle = |j_1,...,j_N\rangle\}$ are the
eigenvectors of $\prod_k \sigma_z^{(k)}$.
The dynamics preserves the qubit configuration $|j\rangle$ and therefore
the evolution operator (\ref{evolution_operator_01}) is
diagonal in the system indices:
\begin{equation} 
\label{evolution_operator_02}
\langle j \,\alpha_{\tiny \textsf E} | U(t) | l \alpha_{\tiny \textsf E}^\prime \rangle 
= \delta_{jl} \,
\langle \alpha_{\tiny \textsf E}| U(t|j) |\alpha^\prime_{\tiny \textsf E} \rangle,
\end{equation}
where $U(t|j)=\langle j | U(t) | j\rangle$ expresses the
conditional evolution operator of the environment alone.
Therefore 
\begin{equation}
(\rho^\prime)_{jl}= 
(\rho)_{jl}\sum_\alpha \langle \alpha_{\tiny \textsf E} | \, U(t|j) \,w\,
U^\dagger (t|l) \,|\alpha_{\tiny \textsf E}\rangle.
\label{eq:spinbosonrho}
\end{equation}
In this basis representation the environment only changes
the off-diagonal elements of $\rho$, while populations are 
preserved.
If the environment is initially in the pure state 
$w_0 \equiv |0\rangle _{\tiny \textsf E}\!\langle 0|$, then 
the equations~(\ref{Ugendeph})-(\ref{generalizedkraus}) are 
recovered.
At any rate, it is sufficient to consider a purification of 
$w$ in an enlarged Hilbert space to write our model as a 
generalized dephasing channel (\ref{Ugendeph}).

For a multimode environment of oscillators
initially at thermal equilibrium, 
$w=\exp(-\beta H_{\tiny \textsf E})$, we obtain
\begin{equation}
\begin{array}{l} 
\label{general_dephasing_factor}
\displaystyle{
\sum_\alpha \langle \alpha_{\tiny \textsf E} | \, U(t|j) 
\,w \,U^{\dagger}(t|l)\,|\alpha_{\tiny \textsf E} \rangle  
=}
\\\quad=\,
\displaystyle{
      \exp\Big[-{\lambda^2} \hskip-5pt \int\limits_0^{\;\;\infty}
      \hskip-3pt\frac{d\omega}{\pi}
      S(\omega)\frac{1-\cos(\omega \tau_p)}{\omega^2}
      \Big| \sum_{k=1}^N(j_k-l_k) 
      e^{i\omega(k-1)\tau}\Big|^2\Big],
}
 \end{array}
\end{equation}
where $S(\omega)$ is the power spectrum of the coupling 
operator $X_{\tiny \textsf E}$.

A central question is if and under which conditions a spin-boson
environment gives a forgetful channel.
Even though we cannot give a rigorous proof, we conjecture on 
physical grounds that an exponential time decay of the bath symmetrized 
autocorrelation function
$C(t) = 1/2 \, \langle X_{\tiny \textsf E}(t)X_{\tiny \textsf E}(0)+ 
X_{\tiny \textsf E}(0)X_{\tiny \textsf E}(t)\rangle$ is a sufficient
condition for forgetfulness.
To support this conjecture, we proof inequality (\ref{eq:forgetful})
in the particular case in which two single channel uses ($N=1$) are separated
by idle times $L\tau$. We consider two qubits ($M=2$ in equation (\ref{eq:forgetful})), 
prepared in a generic input state $\rho$. Then we compute the output
state $\rho^\prime$ from equation (\ref{eq:spinbosonrho}), that is, taking into 
account memory effects, and the output $\tilde{\rho}^\prime$ in the memoryless
limit. We obtain, for a generic monotonic decaying 
autocorrelation function,
\begin{equation}
\Vert \rho^\prime -\tilde{\rho}^\prime \Vert_1 \leq 
4 \lambda^2 g^2 \tau_p^2\, C(L\tau),
\label{eq:spinbosonexp}
\end{equation} 
where the dephasing factor
$g$ is such that $(\rho^\prime)_{01} = g (\rho)_{01}$ and
is readily derived from (\ref{general_dephasing_factor}) 
by letting $N=1$. In particular we consider a Lorentzian power 
spectrum $S(\omega)=2\tau_c/[1+(\omega\tau_c)^2]$. In this case,
the autocorrelation function is $C(\tau)=e^{-\tau/\tau_c}$ and
equation (\ref{eq:spinbosonexp}) is replaced by
\begin{equation}
\Vert \rho^\prime -\tilde{\rho}^\prime \Vert_1 \leq 
4 \lambda^2 g^2 \tau_c^2 (1-e^{-\tau_p/\tau_c})^2 e^{-L\tau/\tau_c}.
\label{eq:spinbosonLorentzian}
\end{equation} 
Inequality (\ref{eq:spinbosonLorentzian}) is the
(\ref{eq:forgetful}) in the particular case $N=1$ and 
$M=2$ (we can set $h=4 \lambda^2 g^2 \tau_c^2$ by noting that 
$(1-e^{-\tau_p/\tau_c})^2<1$). 
We conjecture that (\ref{eq:forgetful}) also holds 
for any $N$ and $M$, since the correlations between blocks of $N$
qubits decay exponentially with the delay time $L\tau$.

A remarkable feature of model (\ref{bosonicbath}) 
is that in the limit 
of perfect memory ($\tau_c\to \infty$) there exists 
for any number $N$ of qubits a decoherence-free subspace
$\mathcal{H}_N^{(f)}$,
corresponding to a qubit train with an equal number of 
$|0\rangle$ and $|1\rangle$ states. Since the dimension 
$d$ of this subspace is such that $\log_2 d \approx N 
-1/2\log_2 N$ at large $N$, then the channel is 
asymptotically noiseless, that is, $Q=1$. 
A coding strategy naturally appears when blocks of
$\bar{N}\gg 1$ qubits can be sent within the memory time scale 
$\tau_c$: 
if the quantum information is encoded in the 
decoherence-protected subspace $\mathcal{H}_{\bar{N}}^{(f)}$
in such a way that the input state $\rho$ is maximally mixed 
within this subspace, then a lower bound for the coherent 
information can be estimated as $I_c(\mathcal{E}_{\bar{N}},\rho)/\bar{N}
\approx \log_2[{\rm dim}(\mathcal{H}_{\bar{N}}^{(f)})]/N
\approx 1-\log_2\bar{N}/(2\bar{N})$.
The memoryless 
dephasing channel instead is recovered in the limit 
$\tau_c\to 0$
and in this case the coherent information is maximized by
the totally unpolarized input states $\rho_{unp}$ and the channel capacity 
$Q=Q_1=1-H(p_0)$, where $p_0=(1+g)/2$.

\begin{figure}
\centering
\includegraphics[width=80mm,height=53mm]{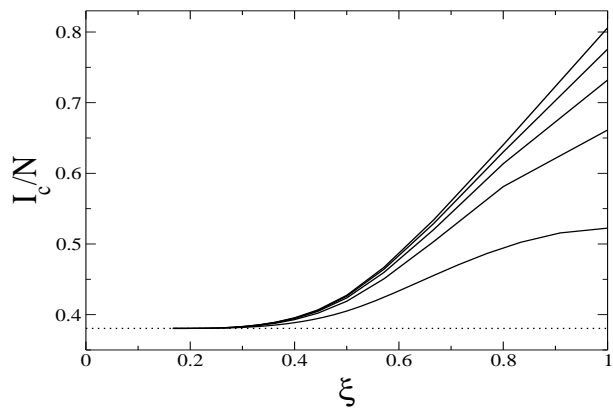}
\caption{Plot of $I_c/N$ as a function of $\xi$,
for the bosonic bath model (\ref{bosonicbath}):
Lorentzian power spectrum, $\lambda=1$,
$\tau_c=1$, $\tau_p=\tau_c$, maximally mixed input state. 
From bottom to top: $N=2,4,6,8,10$.
The dotted line gives the memoryless quantum capacity.}
\label{fig:bosonic}
\end{figure}

Even though we could not compute the channel capacity for
generic values of $\tau,\tau_p$, and $\tau_c$, we show 
in figure~\ref{fig:bosonic} numerical results of the coherent information 
$I_c$ for a Lorentzian power spectrum $S(\omega)$ and for the input state $\rho_{unp}$
as a function of the degree of
memory of the channel, measured by the parameter 
$\xi\equiv\tau_c/(\tau+\tau_c)$. We fix $\tau_c,\tau_p$ and 
vary $\tau$, so that the memoryless and perfect memory 
limits correspond to $\xi\to 0$ ($\tau\to\infty$) and 
$\xi\to 1$ ($\tau\to 0$). 
The curves in figure~\ref{fig:bosonic} show that memory effects
enhance the coherent information $I_c/N$ and that 
$I_c/N$ grows monotonously with $N$. Furthermore, 
these numerical data strongly suggest that $I_c/N$ 
converges, for $N\to\infty$, to a limiting value larger than 
the memoryless capacity $Q_1$. This value 
would provide, assuming the above conjectured forgetfulness
for the model, a lower bound for the quantum capacity. 
Therefore, using the previously mentioned
double blocking strategy, it is possible to 
increase the transmission rate if the quantum information 
is encoded in arbitrarily long blocks,
separated by time intervals larger than $\tau_c$.

\section{Conclusion}
In summary, we have shown that the coherent information in a dephasing
channel with memory is maximized by separable input states, computed 
the quantum capacity $Q$ for a Markov chain noise model and 
suggested a numerical
lower bound for $Q$ in the case of a bosonic bath where memory effects 
decay exponentially with time. 
These results also rely on the concept of forgetfulness, which we prove
for the first model and strongly support on physical grounds for the 
second one.
It would be relevant to further clarify the connection between 
the decay of environment autocorrelation functions and forgetfulness.
It is important to point out that
differently from previous works on quantum memory channels 
\cite{kn:2002-macchiavello-palma-pra},
we have carried out the limit in which the number of channel uses 
$N\to\infty$. 
It would be interesting to investigate
to what extent the results presented in this work could 
be applied to other physically relevant degradable noise models 
such as the amplitude damping channel \cite{giovannettifazio}. 
Another physically relevant question is whether our 
results could be generalized to 
environments with algebraically decaying memory effects, 
which may model typical low-frequency noise in the solid 
state. 
\vspace{0.5cm}

\textit{Note}: After completion of our work we became aware of
a related paper~\cite{pleniovirmani}, 
in which, in particular, the quantum capacity of a Markov 
chain dephasing channel is provided. Their derivation,
not reported in that paper, is based on a 
method different from ours
(S. Virmani and M. Plenio, private communication).
\vspace{0.5cm}

\ack{
We acknowledge fruitful discussions with Vittorio 
Giovannetti, Andrea Mastellone and Rosario Fazio
and anonymous referees for useful remarks. 
G.F. acknowledges P. Salomone and E. Paladino for 
invaluable help.
G.B. acknowledges support from MIUR-PRIN 2005 (2005025204).
A.D. and G.F. acknowledge support from the EU-EuroSQIP (IST-3-015708-IP) and
MIUR-PRIN2005 (2005022977).

\end{document}